\documentstyle [11pt]{article}
\def\D{\Delta}
\def\frac#1#2{{\textstyle{#1\over\vphantom2\smash{\raise.20ex
        \hbox{$\scriptstyle{#2}$}}}}}
\def\co{{\cal O}}
\def\refer#1#2{{{#1}\markboth{REFERENCES}{REFERENCES}}
                \list{{\bf[}\arabic {enumi}{\bf]}}{\settowidth\labelwidth{[#2]}
                        \leftmargin\labelwidth\advance\leftmargin
                        \labelsep\usecounter{enumi}} 
                 }

\def\PL#1#2{Phys. Lett. B\ {\bf{#1}},{#2} }

\def\MPL#1#2{Mod. Phys. Lett. A\ {\bf{#1}},{#2} }

\def\PRL#1#2{Phys. Rev. Lett.\ {\bf{#1}},{#2}}
\def\PRD#1#2{Phys. Rev. D\ {\bf{#1}},{#2}}
\def\RMP#1#2{Rev. Mod. Phys.\ {\bf{#1}},{#2}}
\def\ASS#1#2{Astro. Space Sci.\ {\bf{#1}},{#2}}
\def\SJNP#1#2{Sov. J. Nucl. Phys.\ {\bf{#1}},{#2}}
\def\JETP#1#2{Sov. Phys. JETP\ {\bf{#1}},{#2}}

\def\baselinestretch{1.5}
\def\title#1#2#3#4{
\begin{tabbing}
\= ~                 \hspace{4.5in} ~  \= COLO-HEP-
                                                  #1 \\
\>~                                  \> #2 \\
\end{tabbing}
\vspace{-0.6in}
\begin{center} {\large\bf #3}\\[.1in]
        {\bf #4}\\[.1in] {\it Physics Department, C.B. 390}\\
        {\it University of Colorado, Boulder, CO-80309}\\[.3in]
{\bf ABSTRACT}\\[-1.0in]
       \end{center}
\def\baselinestretch{1.1}\begin{quotation}}
\def\endtitle{\end{quotation} \newpage
}
\begin{document}
\hoffset=-1.0cm
\voffset=-0.5cm
\title
{296}
{~
October,1992
}
{How unique are the MSW parameters ?
}
{P. K. Mohapatra\footnote{e-mail: pramoda@haggis.colorado.edu}}
{
We investigate the effect of
a possible large frozen-in magnetic field in the core of the sun and a neutrino
magnetic moment of the order of $10^{-10}$ Bohr magneton on the MSW parameters.
Such a possibility can completely depolarize the $\nu_{eL}$ resulting in a
factor
of half
of the emitted number. This in combination with MSW can explain the present
experimental
data on solar neutrinos but with a different region of the
parameter space than the regular MSW alone.
In this scenario there is a larger region of parameter space allowed. There is
a large
strip of parameter space corresponding to adiabatic solution and no solution
for large
mixing angle region. The non-adiabatic region still remains as a solution but
shifted
and larger relative to the regular MSW.
}
\endtitle
The substantial difference between the event rate observed at the Homestake
chlorine experiment(HCE)
\cite{hce} and the corresponding prediction of the standard
solar model(SSM)\cite{ssm} has been referred to as
solar neutrino puzzle\cite{jnb}. The ratio of twenty year average of the
number of $\nu_{eL}$
observed in HCE to that predicted by SSM is $0.28 \pm 0.05$.
(Throughout this note we have  combined all the errors in quadrature and
they represent one standard deviation.)
A deviation from the theory has also been seen in
the Kamiokande-II(Kam-II)
experiment\cite{kii} with the corresponding ratio of $0.49 \pm 0.08$. Both
these experiments are sensitive to the high energy neutrinos.
The two Gallium experiments, Gallex and SAGE, on the other hand are sensitive
to the low energy neutrinos constituting the main bulk of all neutrinos coming
from the sun.
Gallex\cite{Gallex}  result is $0.63 \pm 0.16$ of the SSM and
SAGE\cite{SAGE} reported seeing the fraction $0.44 \pm 0.21$ of the expected.
So the combined Gallium result is $0.60 \pm 0.13$ which is higher than the
other
two experiments but far from unity.
Although it is not universally accepted, there has been a
claim that  the chlorine experiment is anti-correlated with the number of
sunspots in one cycle of solar activity\cite{asc}. On the contrary no such
anti-correlation was observed in the Kam-II and Gallium experiments\cite{bmr}.
There exists a long
list of proposals to resolve the puzzle(s). These include non-standard solar
models and exotic particle physics interactions beyond the standard
electro-weak theory. For the various possible explanations the reader is
referred to the book by Bahcall\cite{jnb}. We
will focus our attention only on two  particular explanations, namely
spin(-flavor) precession for neutrino having a large Dirac(Majorana)
magnetic moment and oscillation of neutrino in matter. Oscillation of neutrino
in matter, MSW effect, seems to be the most attractive solution. It is a two
parameter solution with the allowed region of the parameter space determined by
the three known experimental data. We would like to analyze the effect of
complete depolarization of the neutrino due to a possible large frozen-in
magnetic field in the core of the sun on these two parameters.

One possible solution to the solar neutrino puzzle assumes a large
magnetic moment for the neutrino, either Dirac type or Majorana type
\cite{nmm}. For the latter case one
can only have transition magnetic moment. Let us consider the Majorana case
here because this is easier to realize  than the Dirac case
and the constraints coming from astronomical and cosmological arguments do not
apply to this case\cite{rnm}.
Due to the magnetic field in the convective zone of the sun, some $\nu_{eL}$
get spin-flavor precessed into
$\bar{\nu_{\mu}}_{R}$, and hence will miss detection completely in the HCE.
But, $\bar{\nu_{\mu}}_{R}$ will give some
contribution due to neutral current and electromagnetic
interactions in the Kam-II detector which detects Cherenkov radiation
given by the elastically scattered electron.
The fraction of $\nu_{eL}$ remaining as $\nu_{eL}$ is given by:
\begin{equation}
P_{B}(\nu_{eL} \rightarrow \nu_{eL})= 1- \sin^{2}(\mu B L)
\end{equation}
where $\mu$ is the magnetic moment, B is the average magnetic field and L is
the path length of the neutrino inside the magnetic field.
The combination $\mu B L$ has to be very near the first node.
Because the magnetic field in the convective zone varies with the 11-year solar
cycle, this explanation can account for a possible anti-correlation with the
sunspot number\cite{asc}. Because the magnetic field is not expected to be
uniform one should take an average over $\sin^{2}(\mu B L)$
which will give the maximum depletion factor
to be $\frac{1}{2}$. Most important is the fact that the recent Gallium results
very strongly disfavor such a scenario\cite{bmr}.
So simple magnetic spin-flavor precession alone
most probably can
not explain the solar neutrino puzzle. There has been a lot of activity in the
proposal of matter enhanced spin-flavor precession which, like MSW effect for
matter oscillation, can easily account for a large drop in the $\nu_{eL}$
number in an adiabatic situation\cite{rsf}.
There has also been discussion of combining resonant
spin-flavor precession with MSW\cite{msf}.
Although none of these explanations has been
ruled out so far because of only limited experimental data available to
us, we would like to keep an open mind and consider other possibilities.
There are a few things about the above scenario that has to be kept in mind.
The anti-correlation of the Homestake data with the sun-spot number
is not absolutely clear and is very controversial. We need
better statistics to tell one way or the other.
Secondly, unfortunately the internal
magnetic field of the sun is very far from being clearly understood\cite{smf}.
There could exist a frozen-in(primeval) magnetic field as large as 300 MG
near the center of the sun with a magnitude of $2 \pm 1$ MG just beneath
the base of the convective zone.
Even if one has on the average 10 MG over a length of 0.01 $R_{\odot}$,
one will get complete depolarization of the $\nu_{eL}$ for energy $\sim$1 MeV
with a
magnetic moment of $\sim 10^{-10} \mu_{B}$.
It is important to remember that in all the scenarios of the previous kind, one
assumes
that no such large magnetic field exists because that will spoil the whole
motivation of explaining the anti-correlation with solar activity.
We, on the other hand, would like to keep an open mind and analyze
the situation with a very large magnetic field in the center
of the sun which has gotten very little attention.
In this case a neutrino magnetic moment of the order of $10^{-10}
\mu_{B}$, where $\mu_{B}$ is the Bohr magneton, will give rise to complete
depolarization resulting in half the number of each of
$\nu_{eL}$ and $\bar{\nu_{\mu}}_{R}$.

If neutrino has both mixing and magnetic moment which is very natural,
then there will be two separate effects.
Although they will occur simultaneously in the solar
interior, we can consider their effects separately.
The effect of the magnetic field is determined by the quantity
${\int B dx}$. Neither the strength
nor the extent of the frozen-in magnetic field is accurately known. The
magnitude can be any where between zero and 300 MG extending up to
0.1$R_{\odot}$. For the sake of argument, let us assume a
frozen-in magnetic field of strength
10 MG extending up to 0.01$R_{\odot}$.
It is possible for the neutrino with  a
magnetic moment of $\sim 10^{-10} \mu_{B}$ to have complete depolarization.
There will be a suppression factor
multiplying the $\sin ^{2} \mu B L$ term
in (1) because of non-degeneracy of the two states\cite{nmm,rsf,msf}.
This is quantified by an effective $(\Delta m^{2})_{eff}$ which is a
combination of $\Delta m^{2}$ and the effect of matter.
At this point it is worth pointing out that the Majorana case has
advantage over the Dirac case because for the former matter effect can cancel
$\Delta m^{2}$ giving a small effective $\Delta m^{2}$. For the later case the
equivalent $\Delta m^{2}$ is only due to matter effect.
The suppression is negligible if
\begin{equation}
\Delta m^{2} \ll 2 E \mu B
\end{equation}
which is assumed here. This condition puts an upper limit of
$\co(10^{-5} eV^{2})$ on $\Delta m^{2}$
for $B=10 MG$, $\mu=10^{-10} \mu_{B}$ and $E \sim$ 1 MeV and the above
condition is obviously satisfied by higher values of E. Any higher value of
the magnetic field or its extent in the sun will give higher limit for
$\Delta m^{2}$ or lower value of E for which complete depolarization takes
place.
This limit on $\Delta m^{2}$ can be one order of magnitude
higher if we take into account matter effect with density $\sim$ 150
gm/cm$^{-3}$. Because of the high density at the center of the sun, the
neutrino produced by the nuclear reactions is largely the heavy mass
eigenstate. In that case the suppression of spin precession is avoided
altogether. In any case
it is not very difficult to satisfy (2) with a large enough magnetic field
and hence to explain a factor of half depletion in the HCE event rate
after averaging with large number of oscillations for neutrinos of
energy scales relevant to our experiments.

A very simple solution to the solar neutrino puzzle has been proposed\cite{pm1}
using
complete depolarization and short oscillation length vacuum oscillation with
maximal
mixing. With a short wavelength vacuum oscillation and maximal
mixing, the number of $\nu_{eL}$'s reaching the earth can be  reduced by
another factor of half. This can explain the Homestake
chlorine experiment. The difference between the Homestake and the
Kamiokande-II experiments can be attributed to the  contribution
to the Cherenkov radiation in the latter through the neutral current and
electromagnetic interactions of the components which are inert in the
former. The Gallium, especially Gallex, result disfavors this simple minded
solution.
So here we will analyze the effect of complete depolarization
on the MSW solution\cite{pm2}.

The matter enhanced neutrino oscillation, MSW effect, is a very attractive
solution to
the solar neutrino puzzle\cite{msw}. Inside matter $\nu_{eL}$ interacts
with the electrons both via neutral and charged current interactions where as
other kind
of neutrinos interact via only the neutral current interaction. This is the
basis of
virtual enhancement in mixing angle compared to the vacuum mixing angle. So
even with a
small vacuum mixing angle one can explain a large depletion in the total number
of
electron type neutrino after they go through the dense matter in the core of
the sun.
We refer to the many excellent reviews\cite{jnb,kp} on the subject for greater
details.
The basic
effect is parametrized in terms of two parameters, namely, the vacuum mixing
angle and
the difference in mass square. This is represented by what has become the
standard
iso-SNU plots for each experiment.
Because we know the profile of matter density inside the sun, given the
two parameters $\D m^{2}$ and $\sin^{2} (2 \theta)$, we can calculate the
probability for
an $\nu_{eL}$ remaining as an  $\nu_{eL}$ in any particular
detector. The iso-SNU contours have three separate regions. The top horizontal
part
is called the adiabatic section and in this region of parameter space the high
energy
neutrinos are depleted more than the low energy neutrinos. For the diagonal
branch,
known as the non-adiabatic, the energy dependence is reversed. The vertical
section on
the right is the large angle solution which is very close to the vacuum mixing
and
neutrinos of all energy are democratically depleted. Taking the three
experiments into
account and looking at the corresponding three iso-SNU contours we can
determine the
regions of the $\D m^{2}-\sin^{2}(2 \theta)$ that is allowed if the MSW is the
solution to the solar neutrino puzzle\cite{lan}. At present there are two small
distinct regions of the parameter space allowed by the three experiments.
Non-adiabatic
region with $\D m^{2} \sim (0.3- 1.2)\times 10^{-5}$ eV$^{2}$, $\sin^{2}(2
\theta)\sim
(0.4-1.5)\times 10^{-2}$ and the large mixing angle solution with
$\D m^{2} \sim (0.3- 4)\times 10^{-5}$ eV$^{2}$, $\sin^{2}(2 \theta)\sim
(0.5-0.9)$ (Please see ref. \cite{lan} and fig. 4).
As the region of parameter space gets squeezed, the original claim of
MSW solution being robust and valid in a large region of parameter space is
getting
lost. There are also claims in the literature for getting this region of the
parameter space from grand unified theories. Of course, if we knew for sure
that MSW
is {\it {the solution}} to the solar neutrino puzzle we can use all the
experimental
results to determine the two parameters and hence possibly eliminate certain
extensions of the standard model. But right now the situation is exactly the
opposite.
So now the question is, how unique are
the region of parameter space ? We would like to show in this note that in the
presence of a very large magnetic field and with a magnetic moment within the
experimental limits, there is a solution to the solar neutrino puzzle with a
different region of the parameter space.

The neutrino can undergo complete depolarization and MSW oscillation at the
same time
but for all practical purposes we can talk about the two effects separately.
After
complete depolarization we will get half the total amount of $\nu_{eL}$ which
can
undergo MSW oscillation to give the desired result. The difference between this
scenario and regular MSW alone is that here we start from 50\% instead of
100\%.
In figures 1,2 and 3 we have shown the iso-SNU contours for the MSW after
complete
depolarization. It is important to note that these contours are exactly the
same as
the regular MSW\cite{lan} except that their SNU values are
exactly half of the regular MSW.  The shaded region corresponds to the
presently
allowed region of the parameter space for each experiment within 95\%
confidence
level. In fig. 4 we have combined all the three regions and found the common
region of the parameter space. That is the shaded region. We have also
presented
in the same figure the parameter space allowed by regular MSW. Those two
distinct
regions are given in the unshaded regions.  The differences between
the two scenarios can be easily seen.
In the present scenario there is no large mixing angle solution as opposed to
the
regular MSW. On the contrary, there is a large portion of adiabatic region
which
is allowed in this case. This continues over to the non-adiabatic region which
is
shifted and enlarged version of the usual MSW. The implications of this new
scenario
for future experiments is under investigation. Preliminary result indicates
that there
will not be any remarkable difference in the two scenarios in counting
experiments.
But there will be definite difference in the experiments which will look at the
spectrum of neutrinos, for example SNO and BOREXINO. That is because of the
intrinsic nature of the three different regions of MSW parameter space.
In any case, if the Gallium experiment sees
clearly more than 50\% of SSM we can turn the argument around and put limit on
$\mu B$ and if we know $\mu$ from some other source that will
put a limit on B.

We thank K.S. Babu for valuable discussions.
This work was supported in part by the Department of Energy under contract No.
DE-FGO2-91-ER-40672. \\
\newpage
\noindent
\refer{\bf References and Footnotes}{999}
\bibitem{hce} R. Davis, Jr., in {\it Proc. of XIXth Int. Con. on Neutrino
Physics, ``Neutrino'88"}, ed. J. Schneps et. al. (World Scientific Publishing
Co., Singapore, 1989).
\bibitem{ssm} J.N. Bahcall and R.K. Ulrich, \RMP{60}{297(1988)}.
\bibitem{jnb} See, for example, J. N. Bahcall, {\it Neutrino Astrophysics}
                         (Cambridge University Press, Cambridge, England,
1990).
\bibitem{kii} K.S. Hirata et. al., \PRL{63}{16(1989)}; \PRL{65}{1297(1990)};
                                   \PRL{65}{1301(1990)}.
\bibitem{asc} Rowley, J.K. et. al., in {\it American Inst. Phys. Conf. Proc.
              No. 126, p. 1}, ed. M.L. Cherry, W.A. Fowler and K. Lande;
              Davis, R., ICOBAN-86, p. 237, ed. J. Arafune.
\bibitem{Gallex} P. Anselmann et. al. \PL{285}{376(1992)}
\bibitem{SAGE} A. Gavrin, SAGE collaboration, presented at the
               XXVI Int. Conf. on High Energy Physics, Aug. 6-12, 1992, Dallas.
\bibitem{bmr} K.S. Babu, R.N. Mohapatra and I.Z. Rothstein,
\PRD{43}{2265(1991)};
              P. Anselmann et. al. \PL{285}{390(1992)}.
\bibitem{nmm} A. Cisneros, \ASS{10}{2634(1979)};
              K. Fujikawa and R. E. Shrock, \PRL{45}{963(1980)};
              L.B. Okun, \SJNP{44}{546(1986)};
              M.B. Voloshin, M.I. Vysotsky and L.B. Okun, \SJNP{44}{440(1986)};
              M.B. Voloshin and M.I. Vysotsky, \SJNP{44}{544(1986)};
              M.B. Voloshin, M.I. Vysotsky and L.B. Okun, \JETP{64}{446(1986)}.
\bibitem{rnm} R.N. Mohapatra, University of Maryland Report \# PP-90-177;
              K.S. Babu and R.N. Mohapatra, \PRD{42}{3778(1990)};
              K.S. Babu and R.N. Mohapatra, \PRL{64}{1705(1990)}.
\bibitem{rsf} C.-S. Lim and W.J. Marciano, \PRD{37}{1368(1988)};
              E.Kh. Akhmedov, \PL{213}{64(1988)};
              E.Kh. Akhmedov and O.V. Bychuk, \JETP{68}{250(1989)}.
\bibitem{msf} E.Kh. Akhmedov,\JETP{68}{690(1989)};
              C.-S. Lim , M. Mori, Y. Oyama and A. Suzuki, \PL{243}{389(1990)}.
\bibitem{smf} W.A. Dziembowski and P.R. Goode in {\it Inside the Sun,} ed. G.
             Berthomieu and M. Cribier, Kluwer Ac. Publ. and references
therein.
\bibitem{pm1} P.K. Mohapatra, \MPL{6}{3467(1991)};
              P.K. Mohapatra, presented at the workshop on The Many Aspects of
Neutrino
              Physics, Fermilab, Nov 14-17, 1991(unpublished).
\bibitem{pm2} P.K. Mohapatra, presented at the Beyond the Standard Model III,
June 22-24,
              1992, Ottawa (to be published in the proceedings).
\bibitem{msw} S. P. Mikheyev and A. Yu. Smirnov, Sov. J. Nucl. Phys. {\bf 42},
                                                                913(1986);\\
              L. Wolfenstein, \PRD{17}{2369(1979)}.
\bibitem{kp} See, for example, T.K. Kuo and J. Pantaleone, \RMP{61}{937(1989)}.
\bibitem{lan} S.A. Bludman, et. al., University of Pennsylvania Report
\#UPR-0516T;
              J.M. Gelb, W. Kwong and S.P. Rosen, University of Texas Report
              \#UTAPHY-HEP-2
\newpage
\begin{center}
Figure Captions
\end{center}
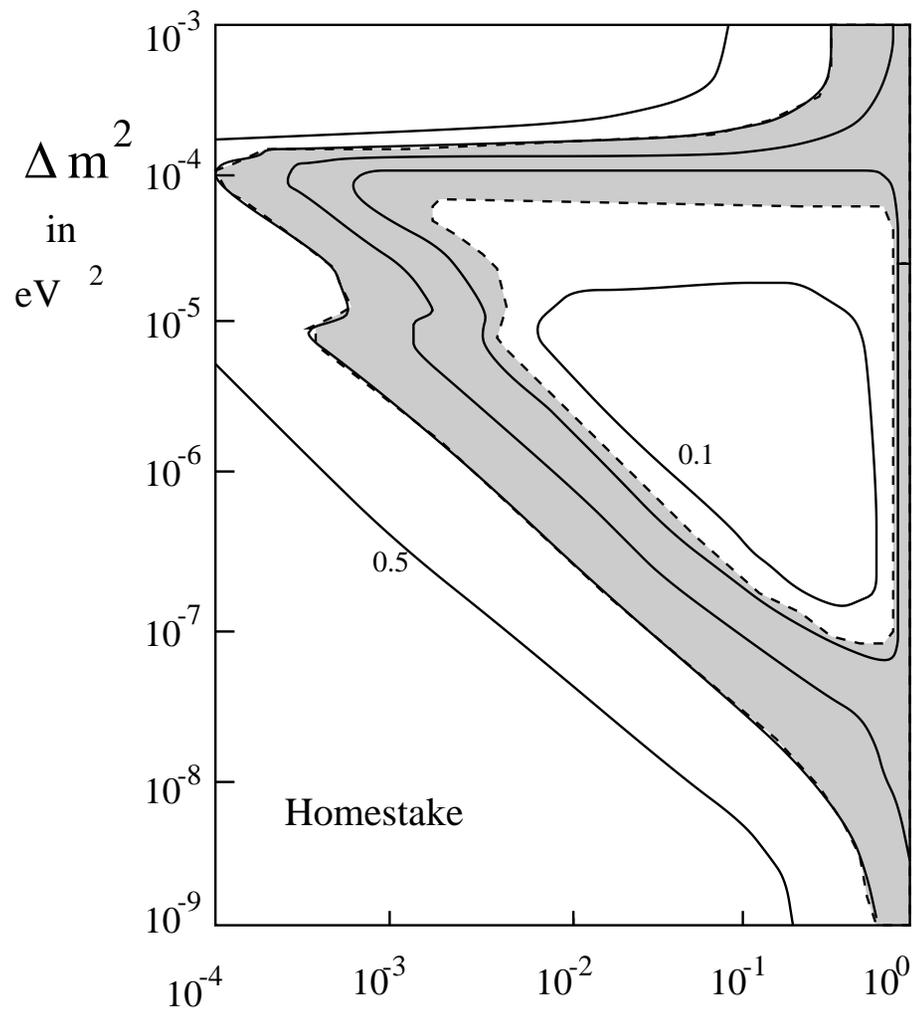
Figure 1. The survival probability contours for $\nu_{eL}$ at Homestake
Chlorine
          experiment for the parameters $\D m^{2}$ and $\sin^{2}(2 \theta)$
after the
          complete depolarization due to the strong magnetic field. The
contours
          represent 0.1, 0.2 ..., 0.5 of the standard solar model of Bahcall
          starting from inside.
          The shaded region represents the allowed region of the parameter
space
          with 95\% confidence level.

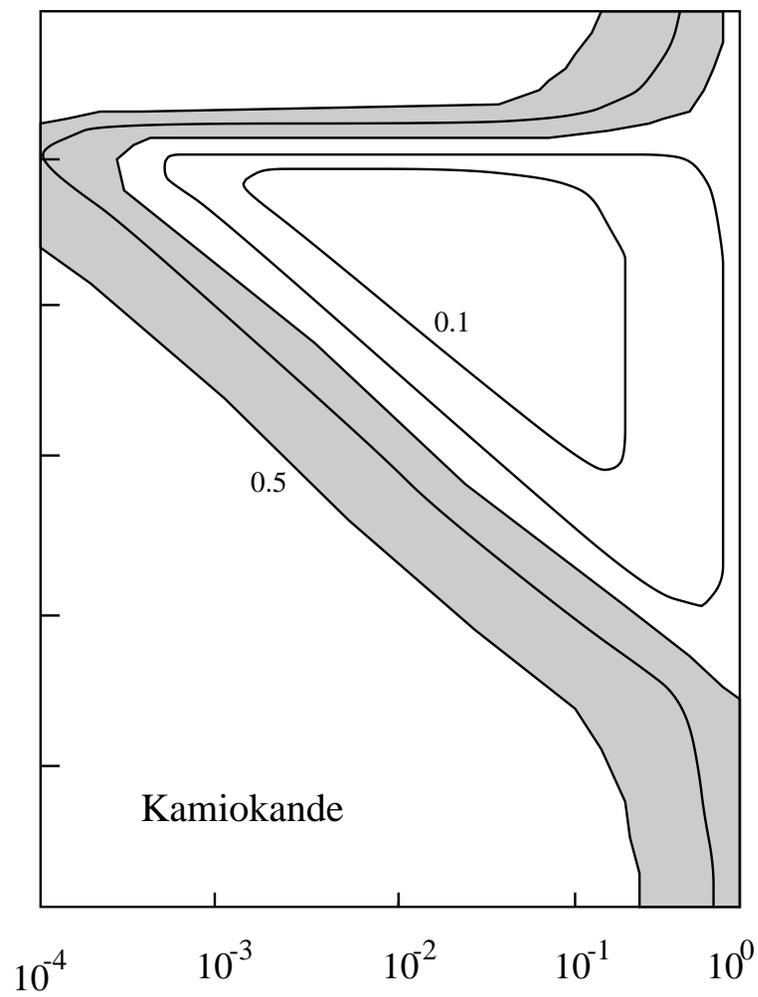
Figure 2. Same as figure 1 but for Kamiokande experiment.

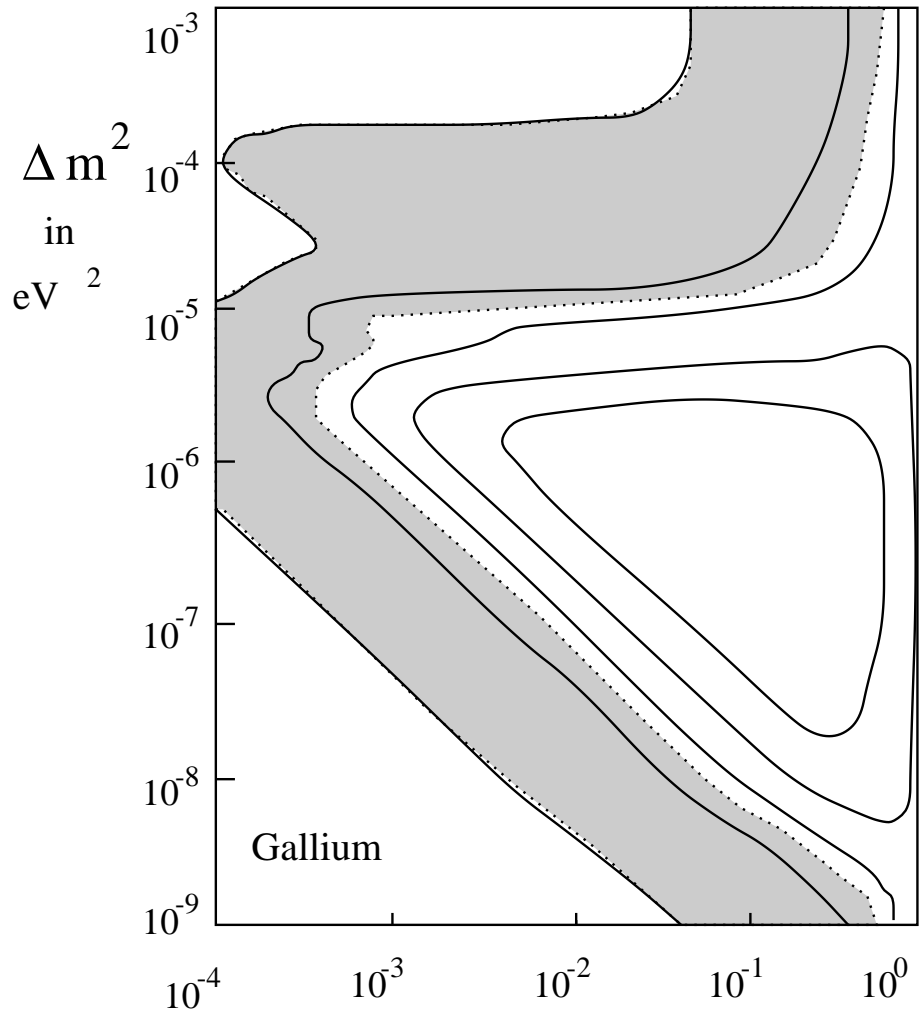
Figure 3. Same as figure 1 but for the combined Gallium experiments.

Figure 4. The region of the parameter space allowed by all the three
experiments. The shaded
          region is for the MSW after complete depolarization while the
unshaded
          region is for MSW alone.
\end{document}